\documentclass[final,5p,times,twocolumn]{elsarticle}
\usepackage{graphicx} 
\usepackage{lineno}
\usepackage{hyperref}
\usepackage{subfigure}
\usepackage{tabularx}

\makeatletter
\renewcommand{\@thesubfigure}{(\alph{subfigure})}
\renewcommand{\p@subfigure}{Figure\space}
\renewcommand{\p@figure}{Figure\space}
\makeatother %

\journal{Nuclear Instruments and Methods A}

\begin{document}

\begin{frontmatter}

\title{HRPPD photosensors for RICH detectors with a high resolution timing capability}

\author[mymainaddress]{A. V.  Lyashenko\corref{mycorrespondingauthor}}
\cortext[mycorrespondingauthor]{Corresponding author}
\ead{alyashenko@incomusa.com}

\author[infn]{J. Agarwala}
\author[tjnaf]{A. Asaturyan}
\author[mymainaddress]{M. Aviles}
\author[bnl]{B. Azmoun}
\author[infn]{C. Chatterjee}
\author[mymainaddress]{S. M. Clarke}
\author[mymainaddress]{S. Cwik}
\author[mymainaddress]{C. J. Hamel}
\author[bnl]{Y. Jin}
\author[bnl]{A. Kiselev}
\author[mymainaddress]{M. J. Minot}
\author[bnl]{B. Page}
\author[mymainaddress]{M. A. Popecki}
\author[bnl]{M. Purschke}
\author[bnl]{S. Stoll}
\author[bnl]{C. Woody}

\address[mymainaddress]{Incom Inc., 294 Southbridge Rd, Charlton MA 01507 USA}
\address[bnl]{Brookhaven National Laboratory, 98 Rochester St, Upton, NY 11973 USA} 
\address[tjnaf]{Thomas Jefferson National Accelerator Facility, 12000 Jefferson Ave, Newport News, VA 23606 USA}
\address[infn]{Istituto Nazionale di Fisica Nucleare, Galleria Padriciano, 99, 34149 Trieste TS, Italy}

\begin{abstract}
Recently, a new version of DC-coupled High Rate Picosecond Photodetectors (DC-HRPPDs) substantially re-designed for use at the Electron-Ion Collider (EIC) has been developed. A first batch of seven “EIC HRPPDs” was manufactured in early 2024. These HRPPDs are DC-coupled photosensors based on Micro-Channel Plates (MCPs) that have an active area of 104 mm by 104 mm, 32 x 32 direct readout pixel array at a pitch of 3.25 mm, peak quantum efficiency in excess of 30\%, exceptionally low dark count rates and timing resolution of 15-20 ps for a single photon detection. As such, these photosensors are very well suited for Ring Imaging CHerenkov (RICH) detectors that can also provide high resolution timing capability, especially in a configuration where a detected charged particle passes through the sensor window, which produces a localized flash containing a few dozens of Cherenkov photons in it.

\end{abstract}
\end{frontmatter}


\section{Introduction}
In many high-energy nuclear and particle physics experiments, Ring Imaging CHerenkov (RICH) detectors are used to identify charged particles over a wide kinematic phase space \cite{reigt2024}, \cite{lhcb2013}, \cite{chatterjee2024}. The choice of photosensor for detecting Cherenkov photons plays a crucial role in ensuring the efficient performance of RICH detectors. In recent years, combining timing information of Cherenkov photons with high gain, high rate capability, and high granularity has become increasingly important. Photon detectors capable of providing time resolutions substantially better than 100 ps for single-photon detection are becoming technologies of significant interest.

\begin{figure}[h]%
\begin{center}%
\subfiguretopcaptrue
\subfigure[][] 
{
    \label{fig:hrppd:front}
    \includegraphics[width=4.1cm]{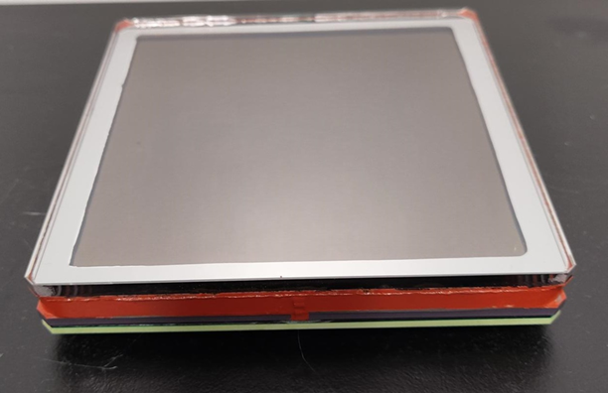}
} \hspace{0cm}
\subfigure[][] 
{
    \label{fig:hrppd:back}
    \includegraphics[width=3.7cm]{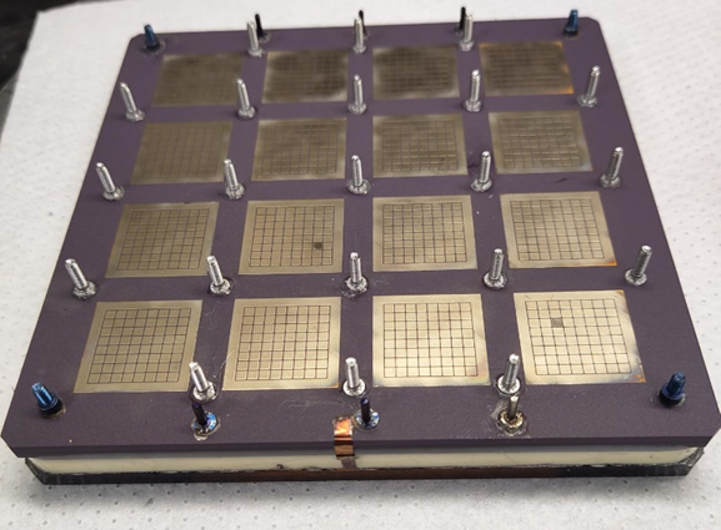}
}\hspace{0cm}
\subfigure[][] 
{
    \label{fig:hrppd:interpos}
    \includegraphics[width=4.1cm]{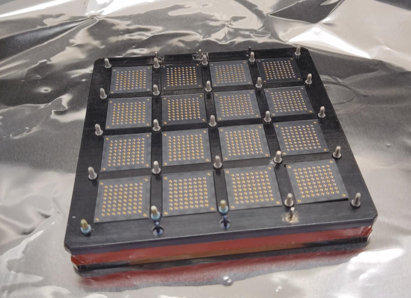}
}\hspace{0cm}
\subfigure[][] 
{
    \label{fig:hrppd:pcb}
    \includegraphics[width=3.7cm]{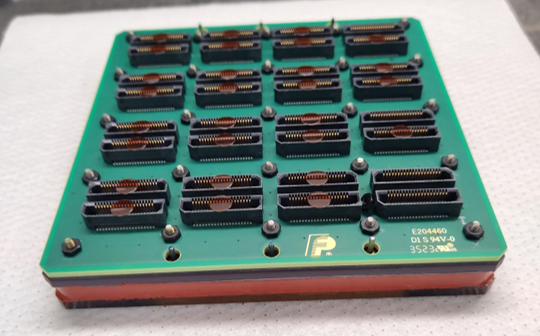}
}\hspace{0cm}
\caption{A photograph of the EIC HRPPD; a) front view, b) back view, c) compression interposers used for signal pickup PCB connectivity and d) test signal pickup PCB attachment}
\label{fig:hrppd}
\end{center}
\end{figure}

A recent version of the DC-coupled High Rate Picosecond Photodetector (DC-HRPPD) developed by Incom Inc. for proximity focused RICH detector at Electron Ion Collider (EIC), fulfills those requirements. These HRPPDs are planar geometry Micro-Channel Plate (MCP) based DC-coupled photosensors featuring an active area of 104 mm by 104 mm, 75\% active area ratio, 32 x 32 array of square pixels spaced at a pitch of 3.25 mm, ALD functionalized 10 $\mu$m pore MCPs and peak quantum efficiency in excess of 30\%. In early 2024, a first batch of seven “EIC HRPPDs” was fabricated to undergo qualification tests for the EIC proximity focused RICH detector. An example of the EIC HRPPD manufactured under this project is shown in \ref{fig:hrppd}.   
The results of a preliminary evaluation of these first DC-HRPPD tiles, including gain and quantum efficiency uniformity, timing resolution, dark count rates, and after-pulse fraction are presented in this article. 

\section{Experimental methods}

\begin{figure}[h]%
\begin{center}%
\subfiguretopcaptrue
\subfigure[][] 
{
    \label{fig:TJNAF}
    \includegraphics[width=4cm]{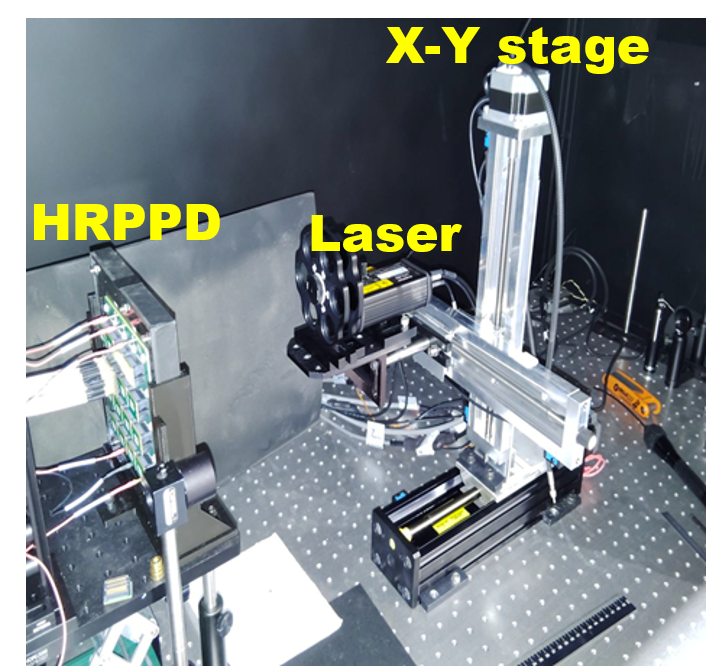}
} \hspace{0cm}
\subfigure[][] 
{
    \label{fig:INCOM}
    \includegraphics[width=4cm]{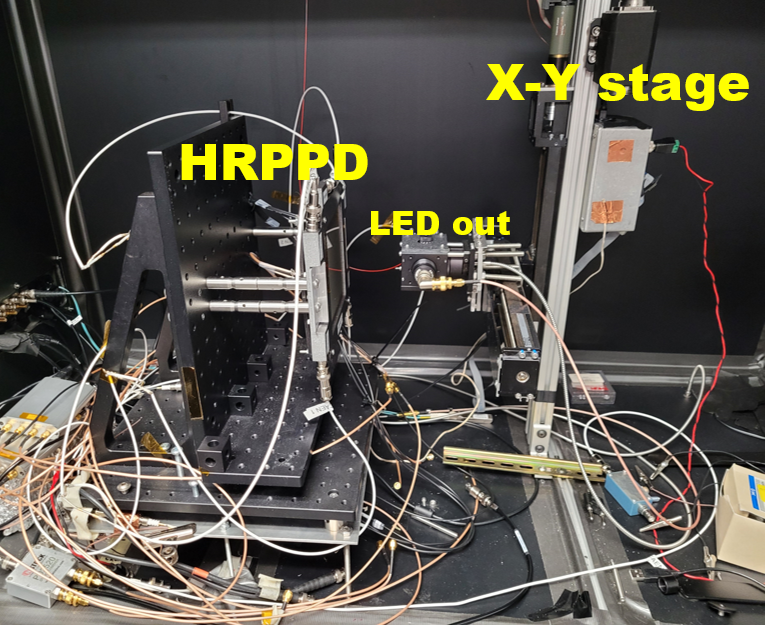}
}\hspace{0cm}
\subfigure[][] 
{
    \label{fig:BNL}
    \includegraphics[width=3cm]{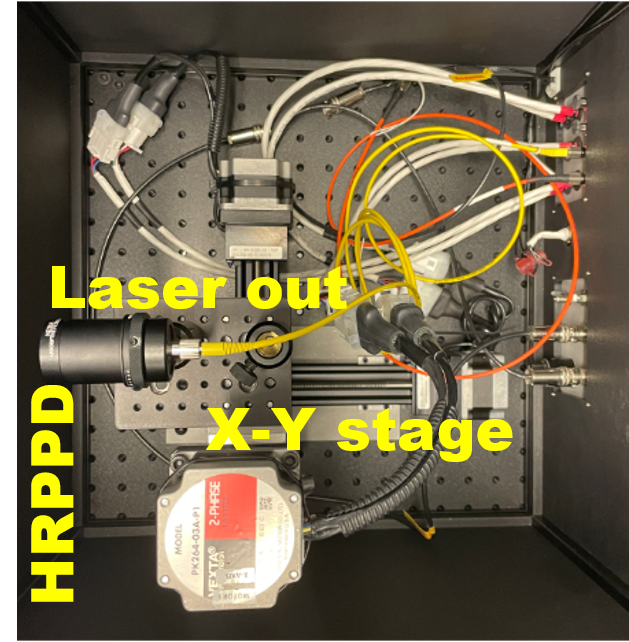}
}
\subfigure[][] 
{
    \label{fig:INFN}
    \includegraphics[width=4cm]{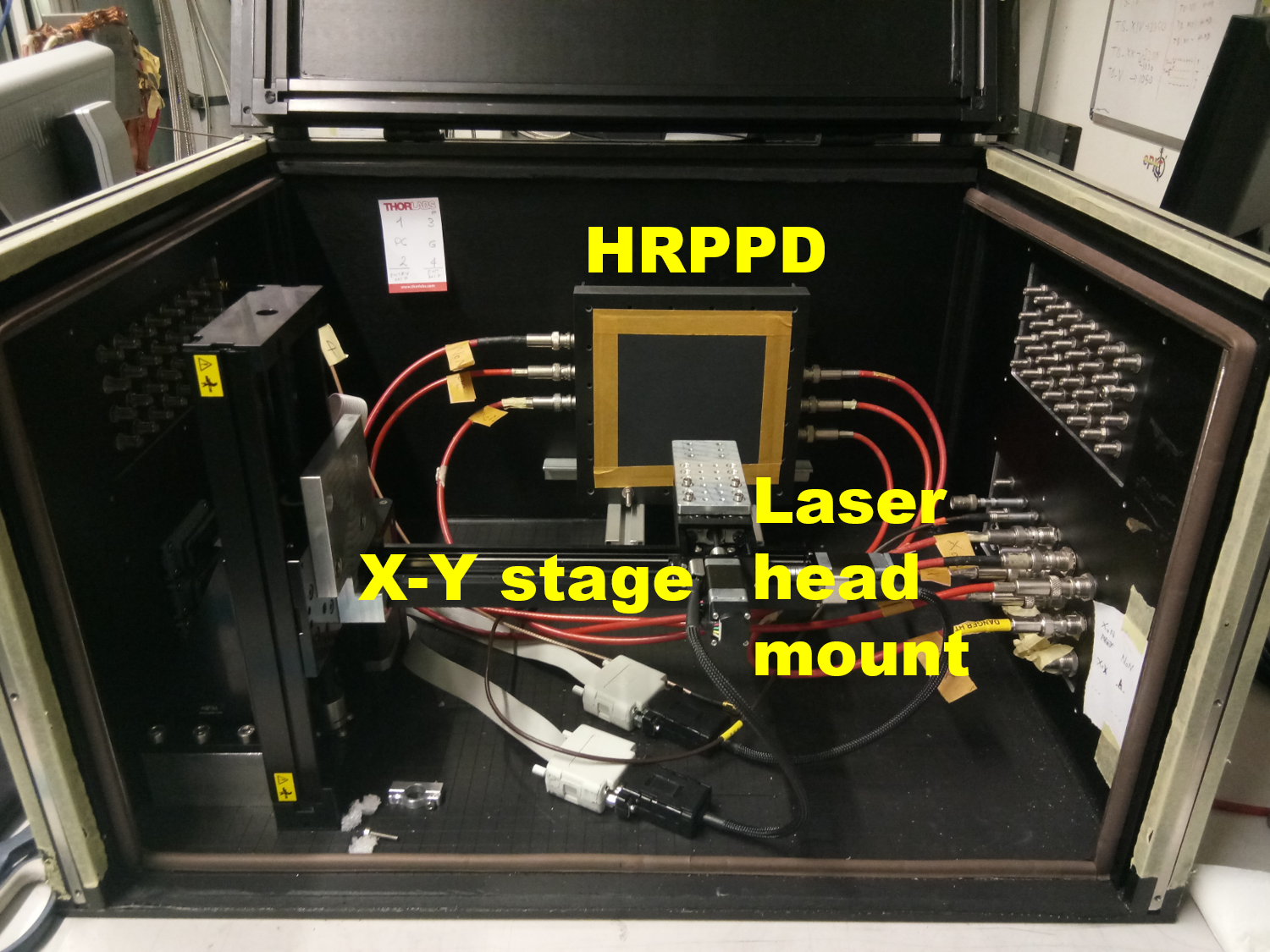}
}
\caption{Photographs of HRPPD test stations at a) TJNAF, b) Incom, c) BNL and d) INFN, Trieste}
\label{fig:setups}
\end{center}
\end{figure}
The HRPPDs were evaluated using dedicated test stations at Thomas Jefferson National Accelerator Facility (TJNAF), Incom Inc., Brookhaven National Laboratory (BNL) and Istituto Nazionale di Fisica Nucleare, Trieste (INFN, Trieste) as shown in \ref{fig:TJNAF}, \ref{fig:INCOM}, \ref{fig:BNL} and \ref{fig:INFN} respectively. These test stations share a similar structure that includes a dark box that accommodates a light source mounted on an X-Y moving stage and a fixture that supports HRPPD. HRPPD signals were picked up using a test PCB connected to the anode pads using custom compression interposers as shown in \ref{fig:hrppd:interpos} and \ref{fig:hrppd:pcb}. The signals were recorded using either a fast scope or appropriate digitizers (e.g. CAEN V1742, featuring DRS4 ASICs). This setup would allow measurements of gain and quantum efficiency uniformity, timing resolution, dark count rates and after-pulse rate. 
\section{Single Photoelectron (SPE) Timing}
\begin{figure}
  \begin{center}
    \includegraphics[width=9cm]{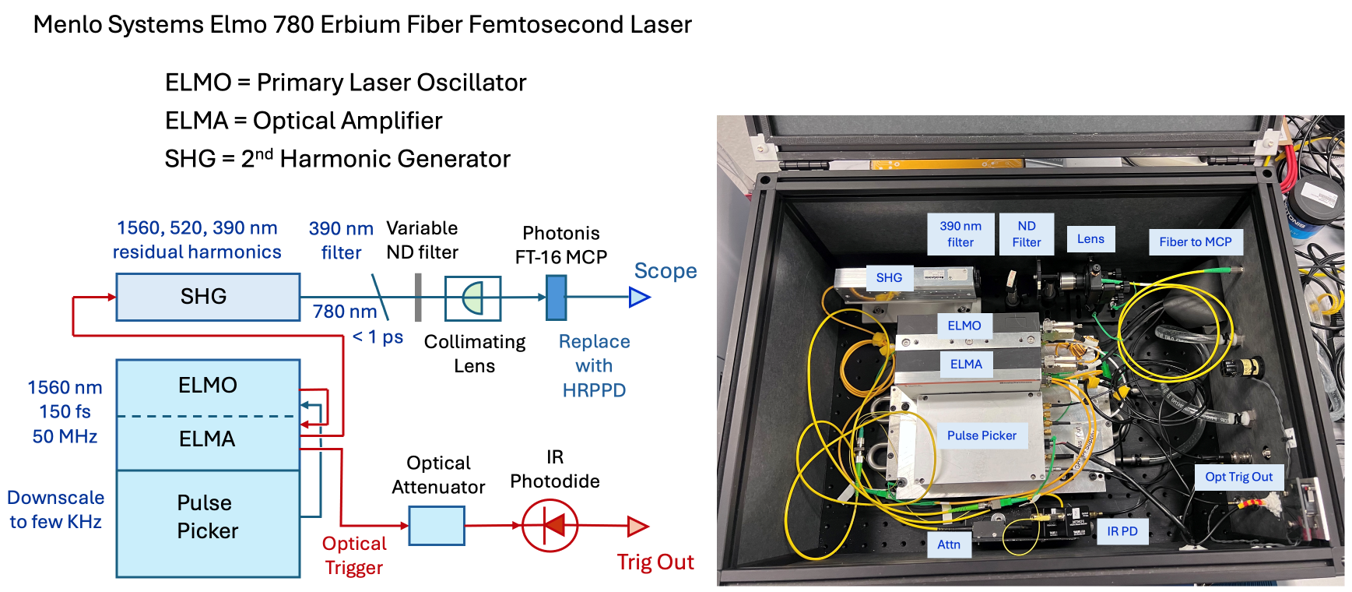}
    \caption{A schematic view (left) and a photograph (right) of an experimental setup for measuring HRPPD timing resolution.}
    \label{fig:timing:station}
  \end{center}
\end{figure}
HRPPD timing was measured using the test setup comprised of the Menlo Systems Model Elmo 780 1560 nm femtosecond laser with a pulse picker followed by a Second Harmonic Generator (SHG) as shown in \ref{fig:timing:station}. The residual third-harmonic 390 nm laser pulse was fed into a test HRPPD through a bandpass filter. The laser beam was focused on a single HRPPD pad center with the intensity tuned down to have $>$ 95\% of empty events ensuring single-photon operation. The signal waveform data were processed with a Tektronix MSO66B scope (50 GS/s, 8 GHz analog bandwidth). To improve data collection efficiency, the scope was triggered on the HRPPD signal at an effective threshold of 5 mV. The timing was then estimated by measuring the arrival time variation in the leading edges of the fast photodiode and HRPPD signals.  

\begin{figure}[h]%
\begin{center}%
\subfiguretopcaptrue
\subfigure[][] 
{
    \label{fig:tts}
    \includegraphics[width=5.5cm]{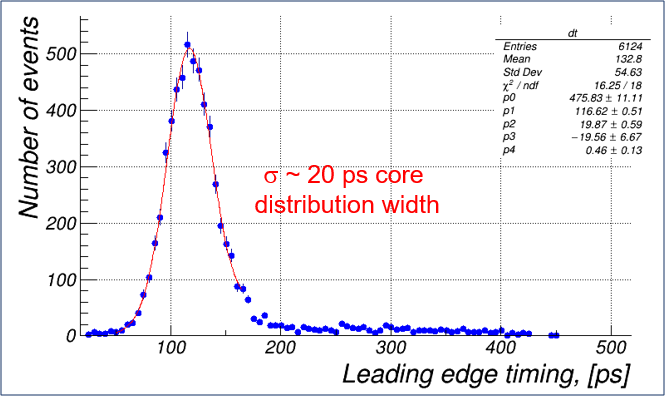}
} \hspace{0cm}
\subfigure[][] 
{
    \label{fig:tts:pc}
    \includegraphics[width=5.5cm]{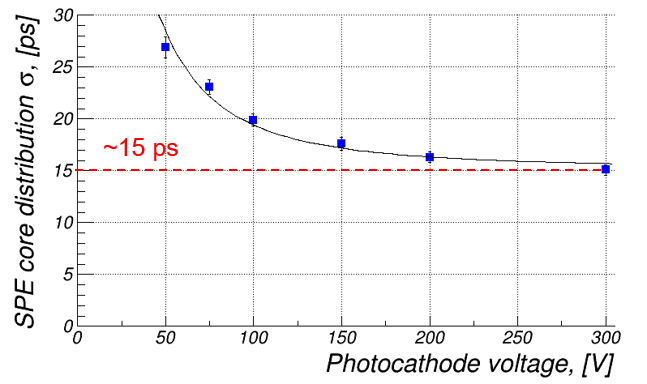}
}\hspace{0cm}
\caption{a) Distribution of the leading edge arrival time differences of the fast photodiode and HRPPD SPE signals; b) SPE timing resolution as a function of HRPPD photocathode gap bias voltage}
\label{fig:timing}
\end{center}
\end{figure}

The distribution of the arrival time differences of the leading edges of the fast photodiode and HRPPD signals is shown in \ref{fig:tts}. The SPE timing resolution was calculated as a standard deviation of the Gaussian function fitted to the data. It was measured to be about 20 ps at MCP bias voltages of 775V and a photocathode gap biased at 100V.      
The SPE timing resolution as a function of the photocathode gap bias voltage is presented in \ref{fig:tts:pc}. The best timing resolution of about 15 ps was achieved at a photocathode gap bias voltage of 300 V. 

\section{Gain}
\begin{figure}[h]%
\begin{center}%
\subfiguretopcaptrue
\subfigure[][] 
{
\label{fig:phd}
\includegraphics[width=5cm]{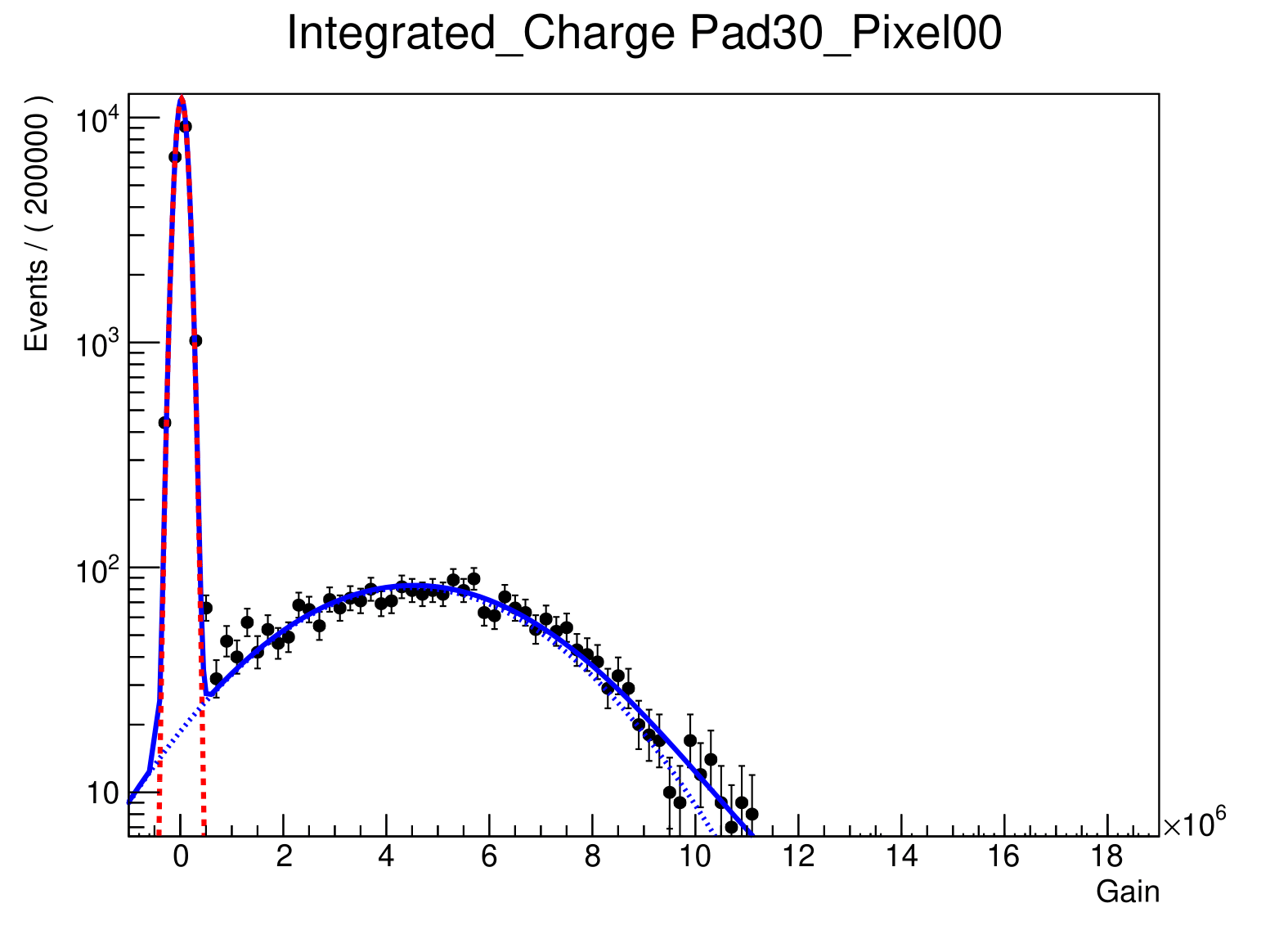}
}\hspace{0cm}
\subfigure[][] 
{
    \label{fig:gain:mcpv}
    \includegraphics[width=6cm]{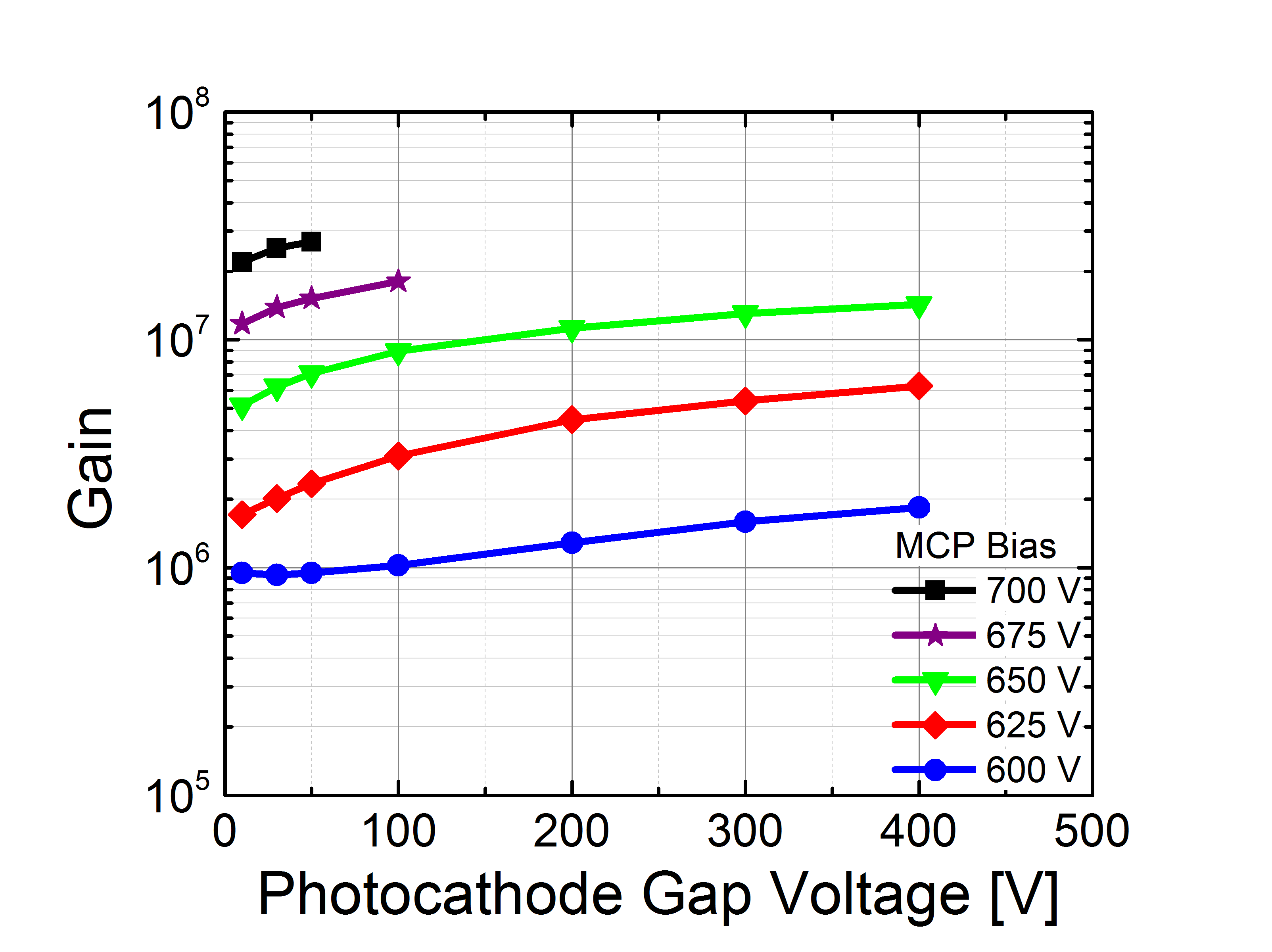}
}\hspace{0cm}
\caption{a) An example of a pulse amplitude distribution recorded at a single pixel of HRPPD; b) HRPPD gain as a function of photocathode gap bias voltage recorded at various MCP bias voltages}
\label{fig:gain}
\end{center}
\end{figure}

\begin{figure}
  \begin{center}
    \includegraphics[width=5cm]{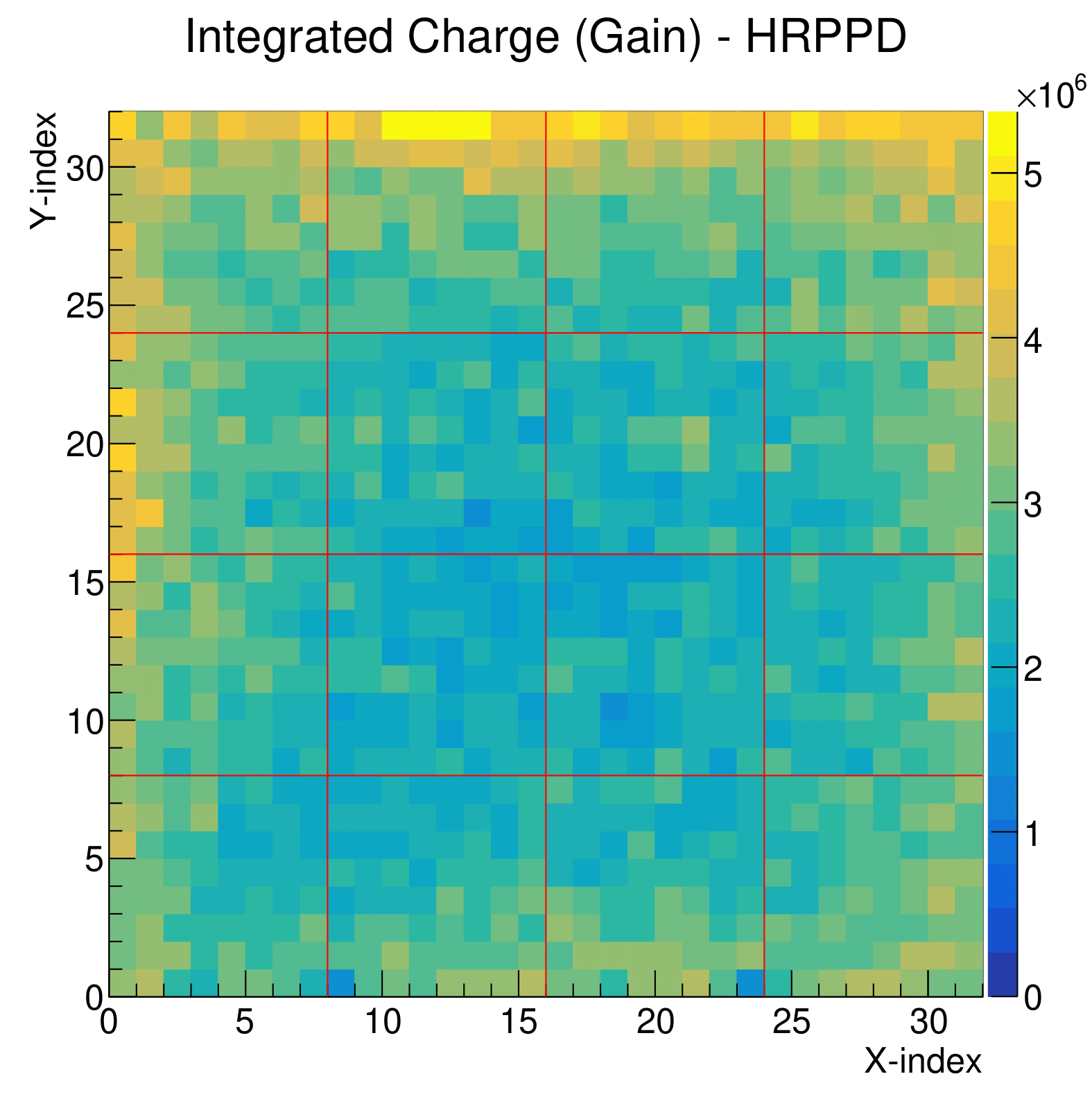}
    \caption{HRPPD gain uniformity map.}
    \label{fig:gain:uni}
  \end{center}
\end{figure}

The HRPPD gain was measured by recording SPE signals from a laser focused on a single pixel. A typical pulse height distribution diagram for these signals is presented in \ref{fig:phd}. The gain was then estimated as the mean value of the pulse height distribution by fitting it with a Gaussian function. To ensure SPE operation, the laser light was attenuated with an appropriate neutral density filter ensuring that the 95\% of laser trigger events were empty.  
The gain as a function of the photocathode gap bias voltage recorded at various MCP voltages is shown in \ref{fig:gain:mcpv}. It is shown that a gain of 10$^7$ can be easily achieved.  
The uniformity of the gain across the area is presented in \ref{fig:gain:uni}. It was visualized by calculating the gain at every pixel of the anode as a mean value of the pulse height distribution of \ref{fig:phd}. The gain in the central region of the tile is fairly uniform and increases by about a factor of two towards the edges of the tile. 

\section{Quantum efficiency (QE)}

\begin{figure}[!ht]%
\begin{center}%
\subfiguretopcaptrue
\subfigure[][] 
{
\label{fig:qe:spec}
\includegraphics[width=5cm]{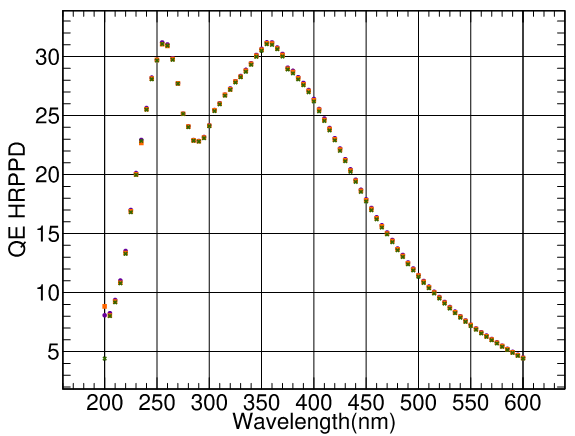}
}\hspace{0cm}
\subfigure[][] 
{
    \label{fig:qe:scan}
    \includegraphics[width=6cm]{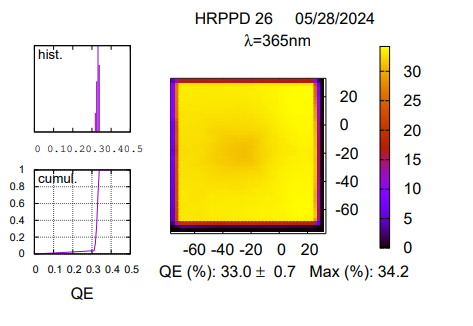}
}\hspace{0cm}
\caption{a) HRPPD QE as a function of wavelength; b) QE uniformity plot}
\label{fig:qe}
\end{center}
\end{figure}

The quantum efficiency was measured by operating the HRPPD in a diode mode when a negative HV was applied to the photocathode while the photocurrent was read out at the entry MCP. To calculate the absolute QE, we compared the HRPPD photocurrent with the photocurrent of a NIST-calibrated photodiode the QE of which is known. The intensity of the light source was monitored by a different photodiode using a beam splitter along the light path. In the spectral QE plot presented in \ref{fig:qe:spec}, it can be seen that the QE has two maxima at around 255 and 355 nm. The maximum QE at 355 nm was measured to be about 32\% for this particular HRPPD.        

The QE uniformity map was measured by a raster scan of an LED in mm-size steps across the HRPPD window. In \ref{fig:qe:scan} it is shown that the QE could be more than 90\% uniform throughout the entire HRPPD active area.     

\section{Dark count rate (DCR)}
\begin{figure}
  \begin{center}
    \includegraphics[width=5.5cm]{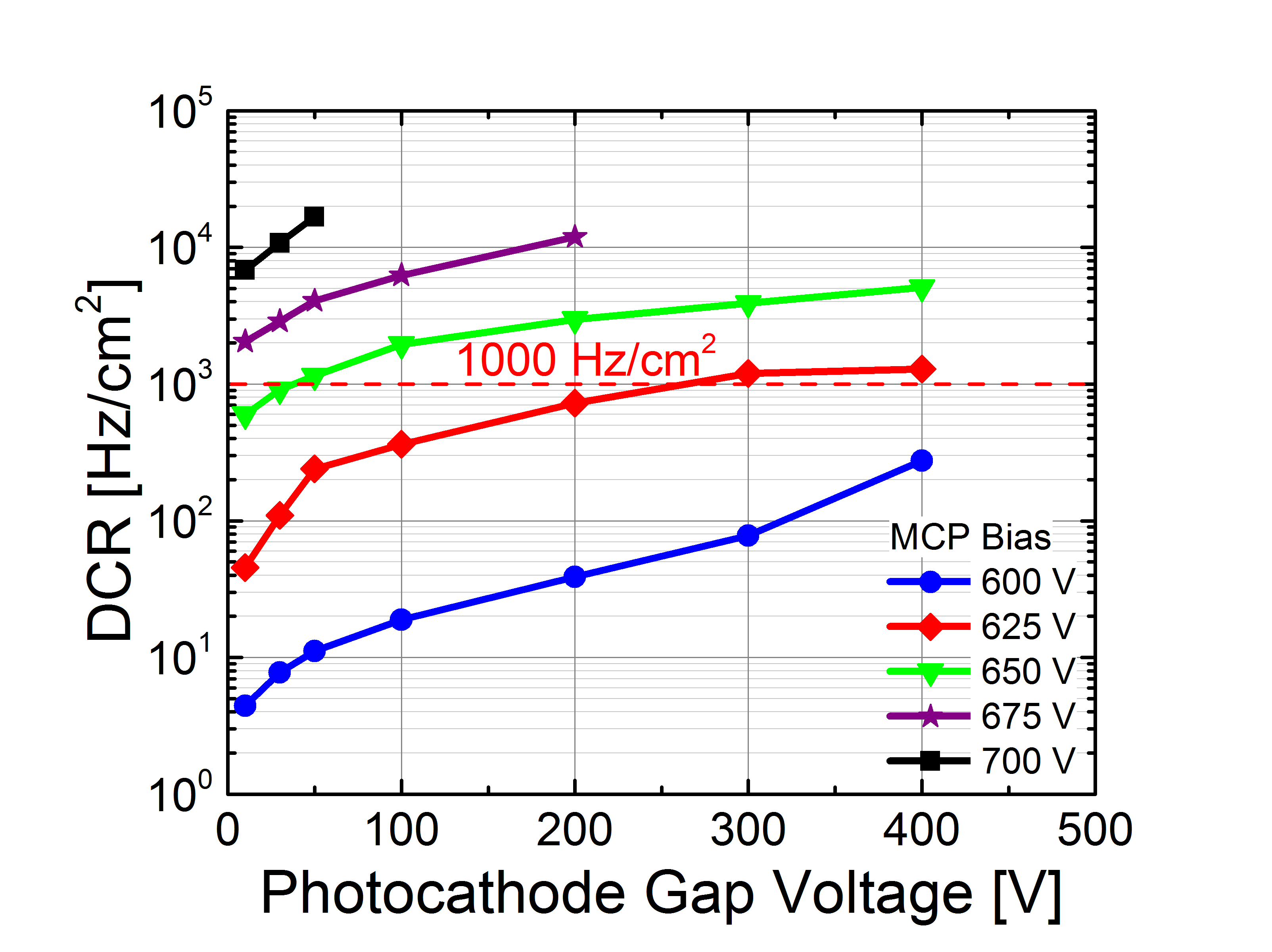}
    \caption{Dark count rate as a function of photocathode gap voltage measured at various MCP voltages.}
    \label{fig:dcr}
  \end{center}
\end{figure}

The DCR was measured by direct readout of HRPPD dark pulses with a 1GHz bandwidth oscilloscope at a threshold of 4mV. The DCR as a function of the photocathode voltage measured at various MCP voltages is shown in \ref{fig:dcr}. The figure also shows a recommended operation points (ROP) region highlighted in red, corresponding to the MCP and photocathode bias voltages for HRPPD operation at a DCR of less than 2 kHz/cm$^2$. For a more thorough quantification of DCR in DC-HRPPDs, one would need to measure the DCR for each of 1024 pixels. These measurements will be performed in the second half of 2025.      

\section{After pulse fraction (APF)}

\begin{figure}[!ht]%
\begin{center}%
\subfiguretopcaptrue
\subfigure[][] 
{
\label{fig:apf:pc}
\includegraphics[width=5.5cm]{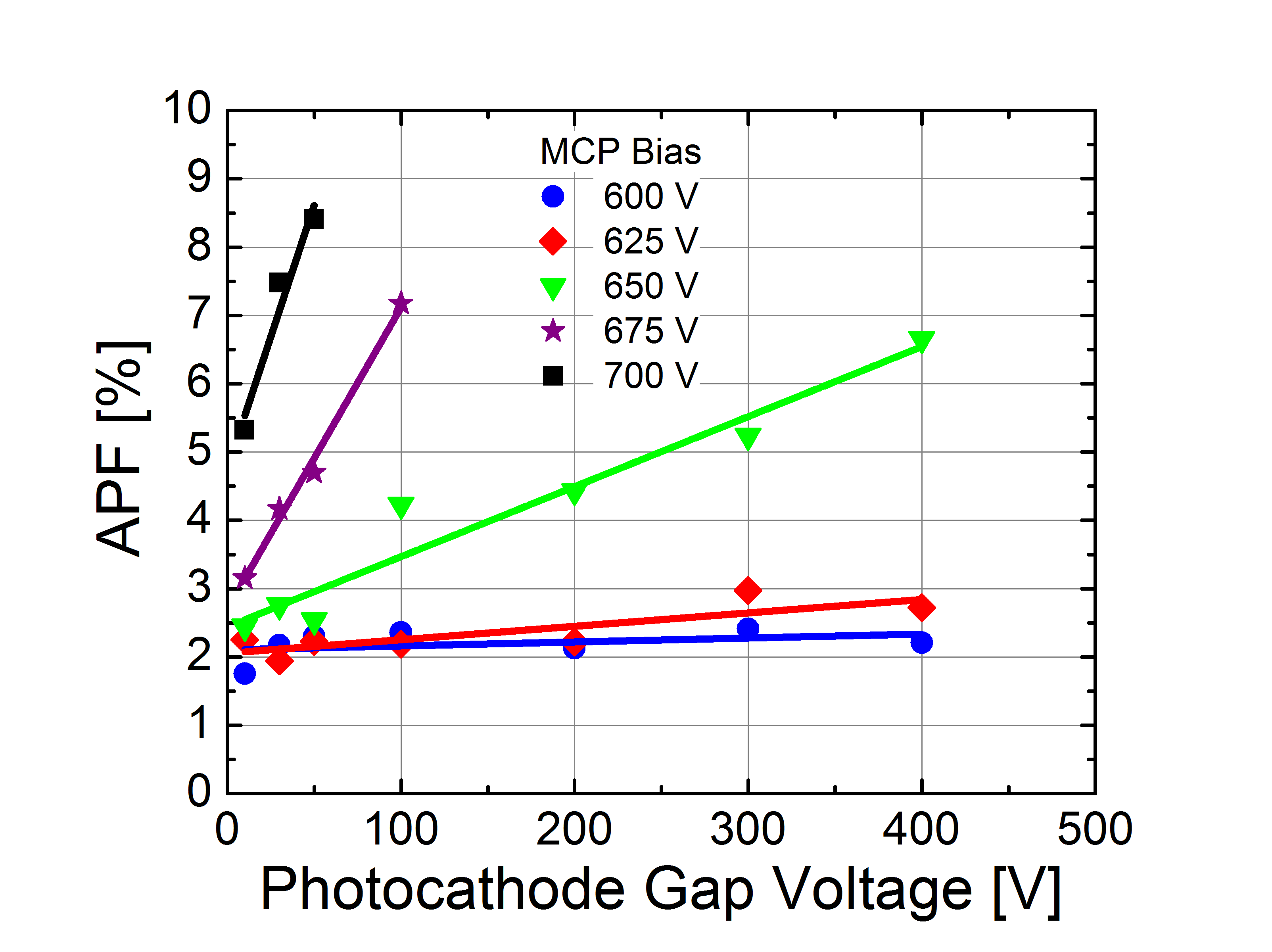}
}\hspace{0cm}
\subfigure[][] 
{
    \label{fig:apf:mcp}
    \includegraphics[width=5.5cm]{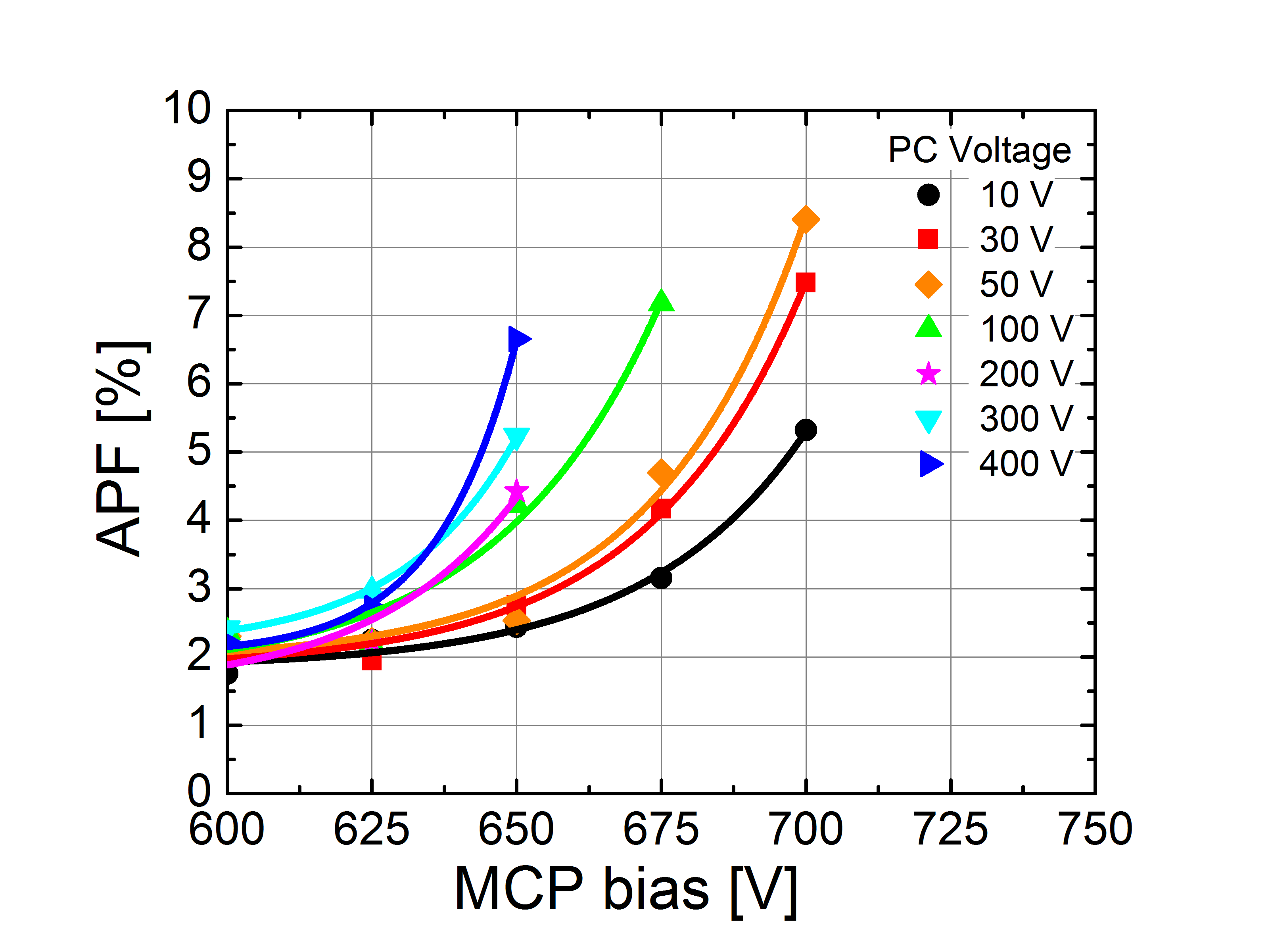}
}\hspace{0cm}
\caption{After-pulse fraction as a function of photocathode gap voltage a) and MCP bias voltage b). Solid lines represent data fitting using a linear a) and an exponential b) function.}
\label{fig:apf}
\end{center}
\end{figure}

Afterpulsing is the mechanism that can adversely affect the lifetime of an MCP-PMT. Trapped molecules in the pores of the MCPs left over from MCP and HRPPD fabrication (such as water, carbonates, nitrogen, etc.) can become ionized and released by the incoming electron cascade. These ions, being positively charged, move up through the MCP against the negative potential applied to the MCP. If this occurs in the entry MCP, the resulting ions impinging on the photocathode may produce some surface damage and a secondary pulse, termed an after-pulse. These pulses occur most commonly 20-100 ns after the initial pulse when the energy from the incoming ion is enough to emit a photoelectron from the photocathode.

The after-pulse fraction, or a percentage of SPE pulses that have an after-pulse, was measured by analyzing waveforms of laser-induced HRPPD signals. It is shown in \ref{fig:apf} as a function of the photocathode gap voltage \ref{fig:apf:pc} and the MCP bias voltage \ref{fig:apf:mcp}. The dependence of the APF on the photocathode voltage can be approximated with a linear function as depicted with solid lines in \ref{fig:apf:pc}. An exponential fit was used for the data points in \ref{fig:apf:mcp} that shows the dependence of the APF on the MCP bias voltages. For a gain of $<$10$^7$, the APF was measured to be $<$4\%. As maintaining a low APF is crucial for the extended life of HRPPDs, the APF could be further reduced by operating the entry MCP at a lower bias voltage while keeping a higher bias at the exit MCP to compensate for a gain loss.      

\section{Conclusions}
In 2024 Incom Inc. manufactured a test batch of seven new design DC-HRPPDs for the EIC collaboration. Their main performance metrics including QE, DCR and APF are summarized in Table\ref{tab}. In the table, the DCR and APF values correspond to the HRPPD gain of 5*10$^6$ and a photocathode gap voltage of 100V. 
\begin{table}[!]
    \centering
    \begin{tabular}{|c||c|c|c|}
    \hline
         HRPPD Serial \# & Mean QE & Dark Rates & APF \\
         \hline
         15 & 33\% & 1.8 kHz/cm$^2$ & 2.8\% \\
         \hline
         16 & 34\% & 0.2 kHz/cm$^2$ & 3\% \\
         \hline
         17 & 35\% & 0.4 kHz/cm$^2$ & 2.5\% \\
         \hline
         23 & 27\% & 0.7 kHz/cm$^2$ & 0.5\% \\
         \hline
         24 & 36\% & 1.3 kHz/cm$^2$ & 0.2\% \\
         \hline
         25 & 27\% & 0.4 kHz/cm$^2$ & 0.3\% \\
         \hline
         26 & 33\% & 1.8 kHz/cm$^2$ & 0.8\% \\
         \hline
    \end{tabular}
    \caption{QE, DCR and APF measured for the 7 HRPPDs manufactured for EIC collaboration. DCR and APF values correspond to the HRPPD gain of 5*10$^6$ and a photocathode gap voltage of 100V}
    \label{tab}
\end{table}
These HRPPDs are characterized by high QE in excess of 30\%, low dark count rates of $<$1.8 kHz/cm$^2$ and low after-pulse fraction of $<$2.8\%. 

The low APF should guarantee a stable long-term HRPPD operation without a significant degradation of the photocathode. This will be verified by the lifetime studies that are currently ongoing at BNL, TJNAF and INF(Trieste). HRPPD performance in magnetic field will also be evaluated in the first half of 2025.

\section{Acknowledgments}
This work was supported in part by the US Department of Energy Division of Nuclear Physics under Contract Nos. DE-SC0012704 and DE-AC02-06CH11357.

\end{document}